\documentclass{PoS}

\usepackage{graphicx}
\usepackage{amsmath}

\newcommand{\fref}[1]{Fig.~\ref{#1}}

\title{On two- and three-point functions of Landau gauge Yang-Mills theory}

\ShortTitle{On two- and three-point functions of Landau gauge Yang-Mills theory}

\author{\speaker{Markus Q. Huber}\\
        Theoriezentrum, Institut f\"ur Kernphysik, TU Darmstadt,
        64289 Darmstadt, Germany\\
        E-mail: \email{markus.huber@physik.tu-darmstadt.de}}

\author{Lorenz von Smekal\\
        Theoriezentrum, Institut f\"ur Kernphysik, TU Darmstadt,
        64289 Darmstadt, Germany\\
        Institut f\"ur Theoretische Physik, Justus-Liebig-Universit\"at,
        35392 Giessen, Germany\\
        E-mail: \email{lorenz.smekal@physik.tu-darmstadt.de}}

\abstract{
We present results for the gluon and ghost propagators and the ghost-gluon vertex obtained from Dyson-Schwinger equations. In the zero temperature case we elaborate on the role of the three-gluon vertex and discuss a model that can capture its qualitative features like its anomalous dimensions and a zero crossing of the dressing function. Our results compare well with lattice data. At non-zero temperature we calculated the ghost propagator which agrees rather well with lattice results already within our simple truncation. These results are used to obtain the temperature dependence of the ghost-gluon vertex. We also explain why the ghost propagator does not react to the phase transition despite its direct coupling to the chromoelectric gluon.
}

\FullConference{31st International Symposium on Lattice Field Theory - LATTICE 2013\\
		July 29 - August 3, 2013\\
		Mainz, Germany}

\begin{document}

\section{Introduction}

Quantum chromodynamics (QCD), one of the building blocks of the standard model of particle physics, is a theory with many interesting features, for example, asymptotic freedom, dynamical mass generation or confinement. The high energy regime is well understood, but at lower energies the strongly coupled nature of QCD impedes progress. Several non-perturbative methods are pursued to tackle the intricacies of low energy QCD. Since every method has its benefits as well as its shortcomings, it is often advantageous to combine them or make comparisons.

Here we follow an approach to QCD that is based on its Green functions. They are the basic building blocks for the calculation of many non-perturbative quantities, for example, the masses and other properties of hadrons, e.g., \cite{Alkofer:2000wg,Cloet:2008re}, and are useful in understanding aspects of confinement, e.g., \cite{Fister:2013bh}. Since Green functions are gauge dependent quantities we have to settle for a specific choice of gauge. A very convenient one from the computational point of view is the Landau gauge: it is covariant, has the minimal number of Green functions, and with hindsight we can say that it is also very amenable for the investigation with functional methods. It can also be implemented straightforwardly on the lattice, even though these implementations are not unique in the infrared due to the Gribov problem as discussed, e.g., in \cite{vonSmekal:2008ws}.

It is noteworthy to say that in the past lattice and functional methods both have made important contributions to the understanding of the non-perturbative behavior of Green functions. For lattice calculations the asymptotic regimes are notoriously difficult due to the finite lattice size and the non-zero lattice spacing. However, in 2007 the situation changed when lattices became large enough \cite{Cucchieri:2007md,Cucchieri:2008fc,Sternbeck:2007ug,Bogolubsky:2009dc} to proceed far enough into the infrared (IR) to challenge the so-called scaling solution found with functional equations \cite{vonSmekal:1997vx,vonSmekal:1997is,Pawlowski:2003hq}. The solution found on the lattice became known as decoupling or massive solution. Soon thereafter it could also be reproduced with functional methods \cite{Boucaud:2008ji,Aguilar:2008xm,Fischer:2008uz,Alkofer:2008jy} and also by an extension of the Gribov-Zwanziger action \cite{Dudal:2008sp,Dudal:2011gd,Gracey:2010cg}. Indeed it turned out that with the latter one can obtain a family of decoupling solutions \cite{Boucaud:2008ji,Fischer:2008uz}. It is currently not settled what the source of this ambiguity is. Possible explanations include Gribov copy effects \cite{Cucchieri:1997dx,Bogolubsky:2005wf,Maas:2009se,Maas:2011se,Sternbeck:2012mf}. While this is certainly an interesting topic by itself, it must be stressed that this effects only the deep IR and physical quantities seem unaffected by this, for example, the phase transition temperatures \cite{Fischer:2009gk}. Hence, for physical applications, it is more important to focus on the mid-momentum regime which is the focus of this work.

An obvious advantage of functional equations is the absence of the notorious sign problem of lattice QCD at non-zero density. Albeit cumbersome, calculations in this regime are possible; for a few examples see \cite{Braun:2009gm,Fischer:2011mz,Fischer:2012vc,Muller:2013pya,Muller:2013tya,Fischer:2013eca}. The issue functional equations have to deal with are the required truncations of the infinite tower of equations. Thus, where possible, comparisons with results from other methods can serve as useful benchmarks to judge the reliability of the employed truncations and can provide an idea of how far to trust the results.

Since truncations of functional equations are usually motivated by the asymptotic behavior it is a priori not clear if extensions of the truncation improve the results quantitatively. The only way to find out is to actually carry out the calculation. If the answer is affirmative, we are rewarded by a host of possible future applications ranging from the calculation of quantities that are currently too costly on the lattice (e.g. four-point functions) to calculations at non-zero density.

Here we present one step beyond the truncation scheme that was the state of the art for 15 years by including the ghost-gluon vertex into the set of equations to be solved \cite{Huber:2012kd}. The ghost-gluon vertex always played a central role for the truncation of functional equations since it becomes momentum independent for asymptotically low and high momenta \cite{Lerche:2002ep}. The deviation from the tree-level expression is only minor \cite{Huber:2012kd,Schleifenbaum:2004id,Cucchieri:2008qm,Aguilar:2013xqa,Pelaez:2013cpa} and induces only very small quantitative changes in the gluon propagator. The main quantitative changes necessarily can then be expected from the three-gluon vertex and the two-loop diagrams (which also contain the former). In this work we employ a model for the former whose form is motivated by analytic and lattice results \cite{Cucchieri:2008qm} and that effectively includes contributions from the two-loop diagrams.
All presented calculations used the programs \textit{DoFun} \cite{Alkofer:2008nt,Huber:2011qr} and \textit{CrasyDSE} \cite{Huber:2011xc}.

\section{Zero temperature}

The three-gluon vertex plays an important part in the gluon DSE. Its detailed form directly impacts the gluon dressing in the mid-momentum regime. In ref.~\cite{Huber:2012kd} we proposed a model as an extension of the model of ref.~\cite{Fischer:2002eq} that has the following properties: 1.) It is Bose symmetric. 2.) It has the correct anomalous dimension. 3.) It features a zero crossing. The last property was first observed in three and two dimensions on the lattice \cite{Cucchieri:2008qm,Cucchieri:2006tf,Maas:2007uv} and later on confirmed with DSEs for two, three and four dimensions \cite{Huber:2012kd,Campagnari:2010wc,Huber:2012zj}. On the lattice no points with negative values have been observed in four dimensions \cite{Cucchieri:2008qm}, but the position of the zero crossing extracted from a leading order DSE calculation is at lower momenta than currently available \cite{Huber:2012kd}. The model has the form $D^{A^3}(x,z,y)=D^{A^3,IR}(x,y,z)+D^{A^3,UV}(x,y,z)$ with
\begin{align}\label{eq:3g}
D^{A^3,UV}(x,y,z)&=G\left(\frac{x+y+z}{2}\right)^{\alpha}Z\left(\frac{x+y+z}{2}\right)^{\beta},\\
 D^{A^3,IR}(x,y,z)&=h_{IR} \,G(x+y+z)^{3}(f^{3g}(x)f^{3g}(y)f^{3g}(z))^4,
\end{align}
where $ f^{3g}(x):=\Lambda^2_{3g}/(\Lambda_{3g}^2+x)$ serves to damp the IR part for higher momenta and $Z(x)$ and $G(x)$ are the gluon and ghost propagator dressing functions. $x$, $y$ and $z$ are squared momenta. $h_{IR}$ is chosen as $-1$ and $\Lambda_{3g}$ is a scale parameter that determines the position of the zero crossing. While $D^{A^3,IR}(x,y,z)$ describes the IR behavior of the vertex only roughly, it turns out that $D^{A^3,UV}(x,y,z)$ is quite a reliable approximation in the UV. The exponents $\alpha$ and $\beta$ are chosen such as to reproduce the correct anomalous UV dimension of the vertex and to render $D^{A^3,UV}(x,y,z)$ constant in the IR, viz., $\alpha=-17/9$ and $\beta=0$ for decoupling.

\begin{figure}[tb]
 \begin{center}
  \includegraphics[width=0.49\textwidth]{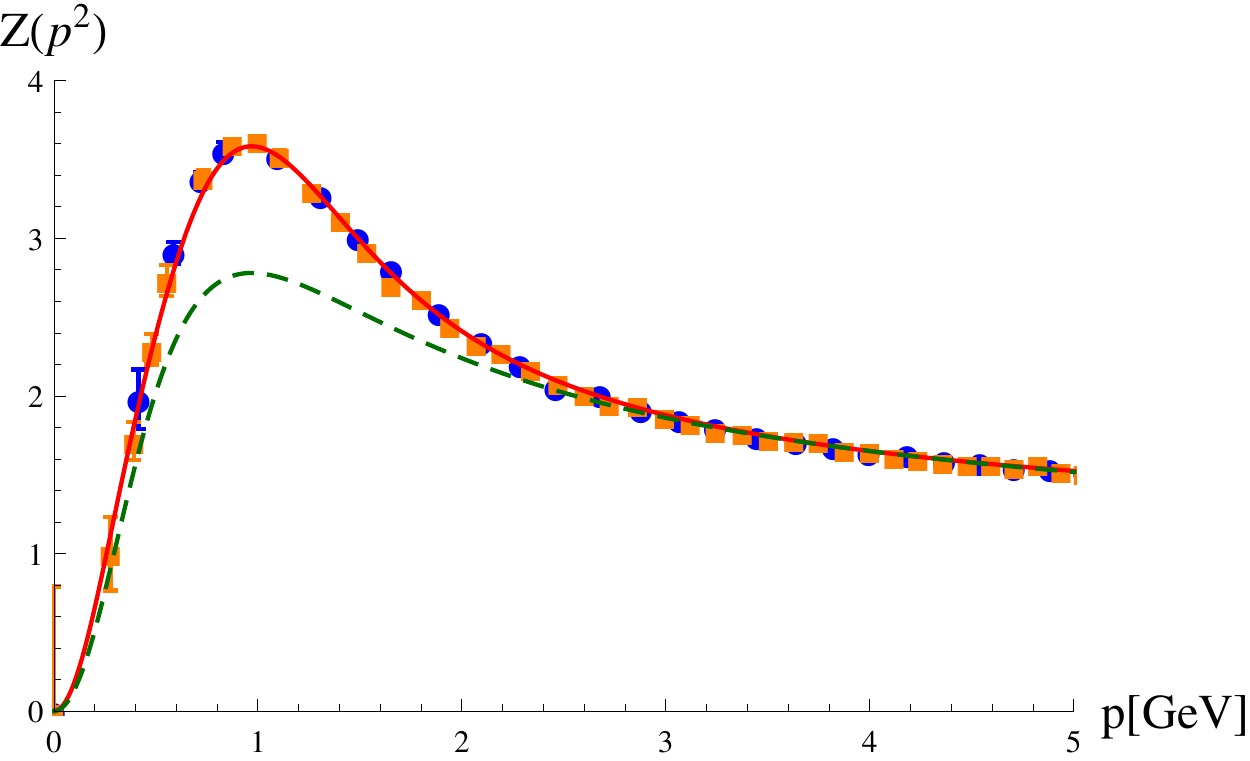}
  \includegraphics[width=0.49\textwidth]{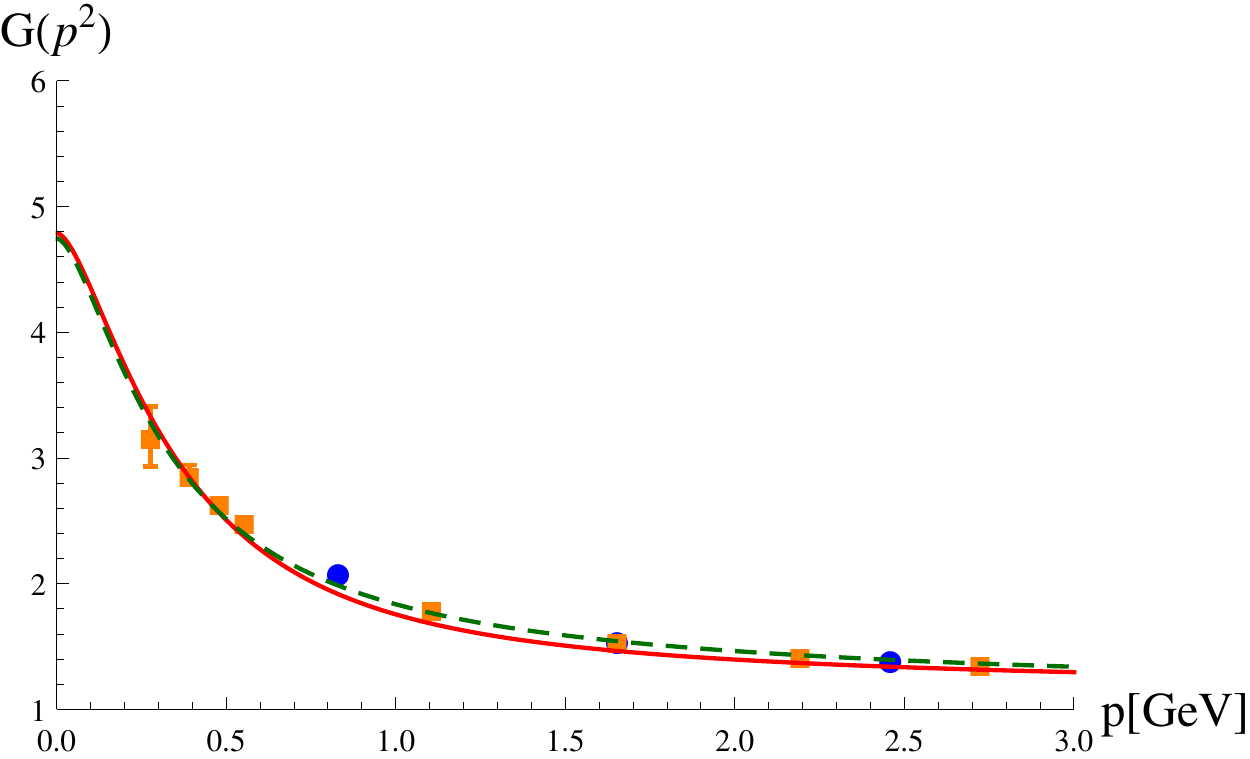}
  \caption{\label{fig:props}The gluon and ghost dressing functions $Z(p^2)$ and $G(p^2)$ in comparison with lattice data \cite{Sternbeck:2007ug}. The red/continuous lines represent the results with a dynamic ghost-gluon vertex and the optimized effective three-gluon vertex, the green/dashed lines a reference calculation with a bare ghost-gluon vertex and the three-gluon vertex of ref.~\cite{Fischer:2002eq}.}
 \end{center}
\end{figure}

\begin{figure}[tb]
\includegraphics[width=0.5\textwidth]{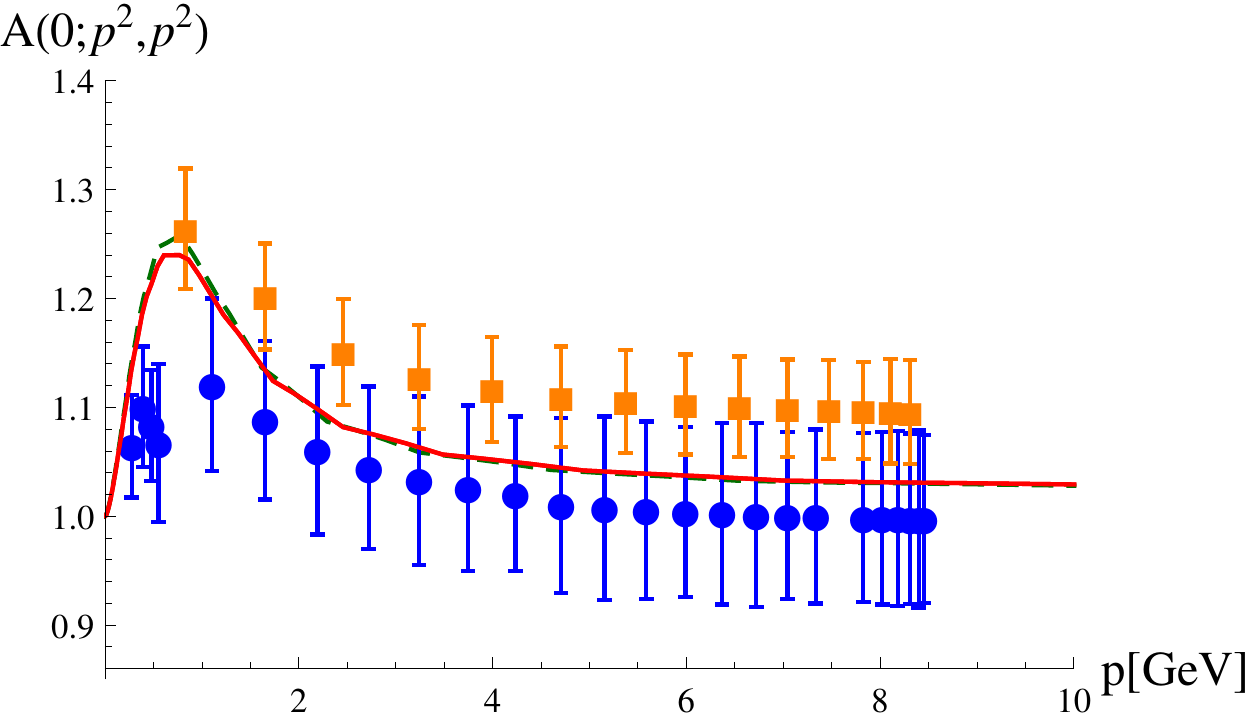}
\includegraphics[width=0.5\textwidth]{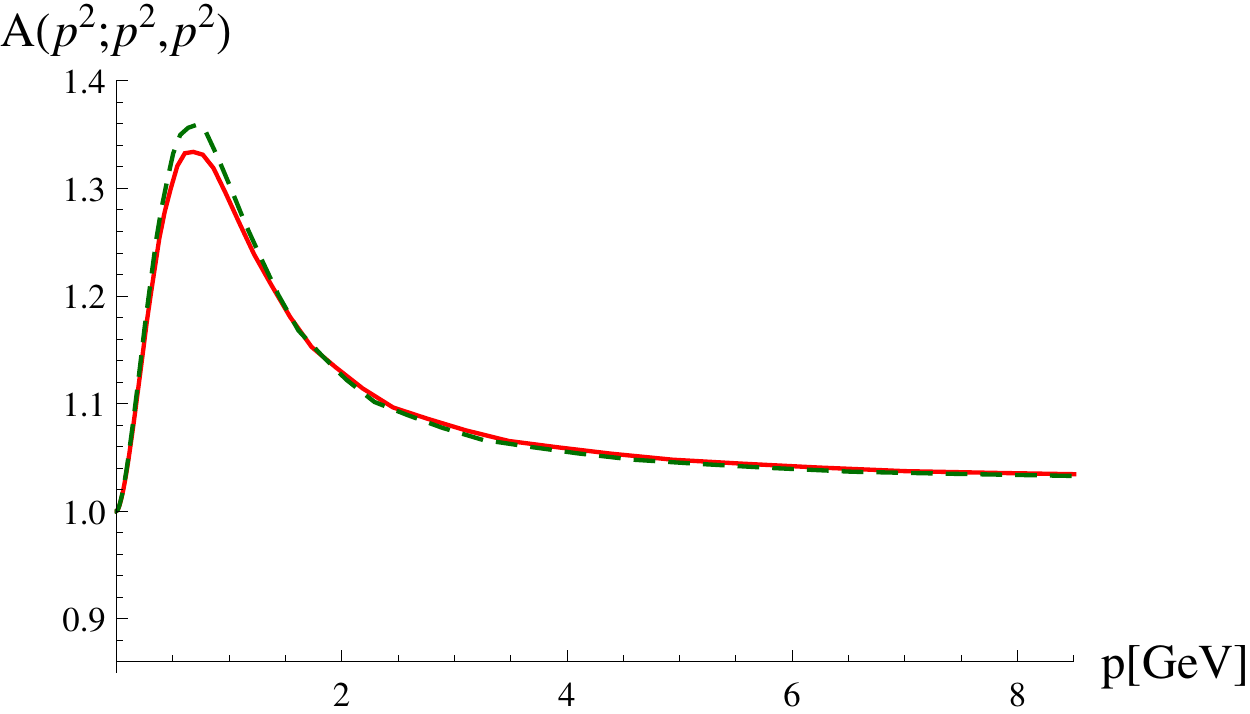}
\caption{\label{fig:ghg}Ghost-gluon vertex dressing function at vanishing gluon momentum (left) and symmetric point (right). The two lines correspond to results from different three-gluon vertex models. Lattice date from \cite{Ilgenfritz:2006he}.}
\end{figure}

The employed model can be used to effectively include two-loop contributions by adjusting the zero crossing. We use this freedom in the present model to optimize our results in the mid-momentum regime. Fig.~\ref{fig:3g_gh_T} compares the optimized effective three-gluon vertex with lattice results and a leading order DSE calculation using the final propagators and ghost-gluon vertex. Results for the system of propagators and ghost-gluon vertex are shown in figs.~\ref{fig:props} and \ref{fig:ghg}. The propagator dressing functions are in good agreement with lattice results. For comparison we also show results from a propagator-only calculation with the three-gluon vertex of ref.~\cite{Fischer:2002eq}. The quality of our results is comparable to those from the functional renormalization group, see \cite{Fischer:2008uz} for the propagators and \cite{Fister:2011uw} for the ghost-gluon vertex.

\section{Non-zero temperature}

\begin{figure}[tb]
\begin{center}
\includegraphics[width=0.49\textwidth]{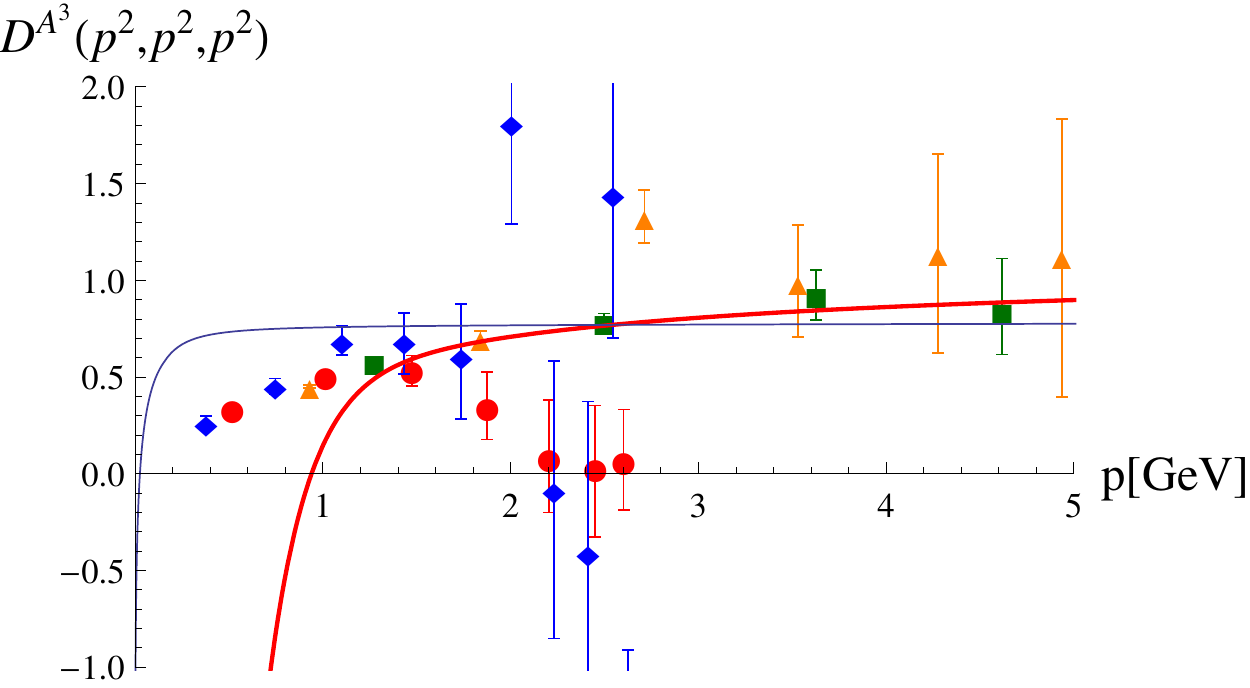}
\includegraphics[width=0.49\textwidth]{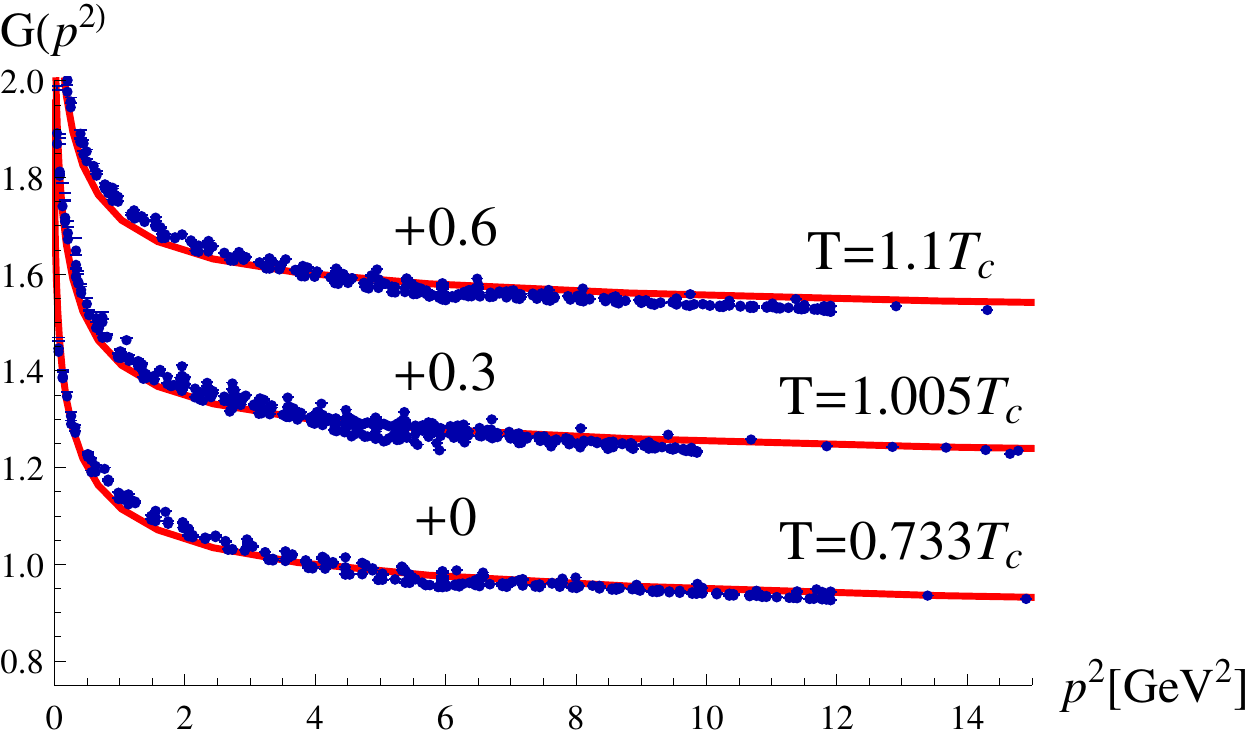}
\caption{\label{fig:3g_gh_T}Left: Lattice results from ref.~\cite{Cucchieri:2008qm} (dots) in comparison with the employed model (thick/red line) and a ghost-triangle-only calculation (thin line) of the three-gluon vertex at the symmetric point. Right: The ghost dressing function at $T=0.733\,T_c$, $1.005\,T_c$ and $1.1\,T_c$ compared to lattice data from ref.~\cite{Fischer:2010fx,Maas:2011ez}. The data was shifted as indicated to distinguish the different temperatures.}
\end{center}
\end{figure}

\begin{figure}[bt]
\begin{center}
\includegraphics[width=0.49\textwidth]{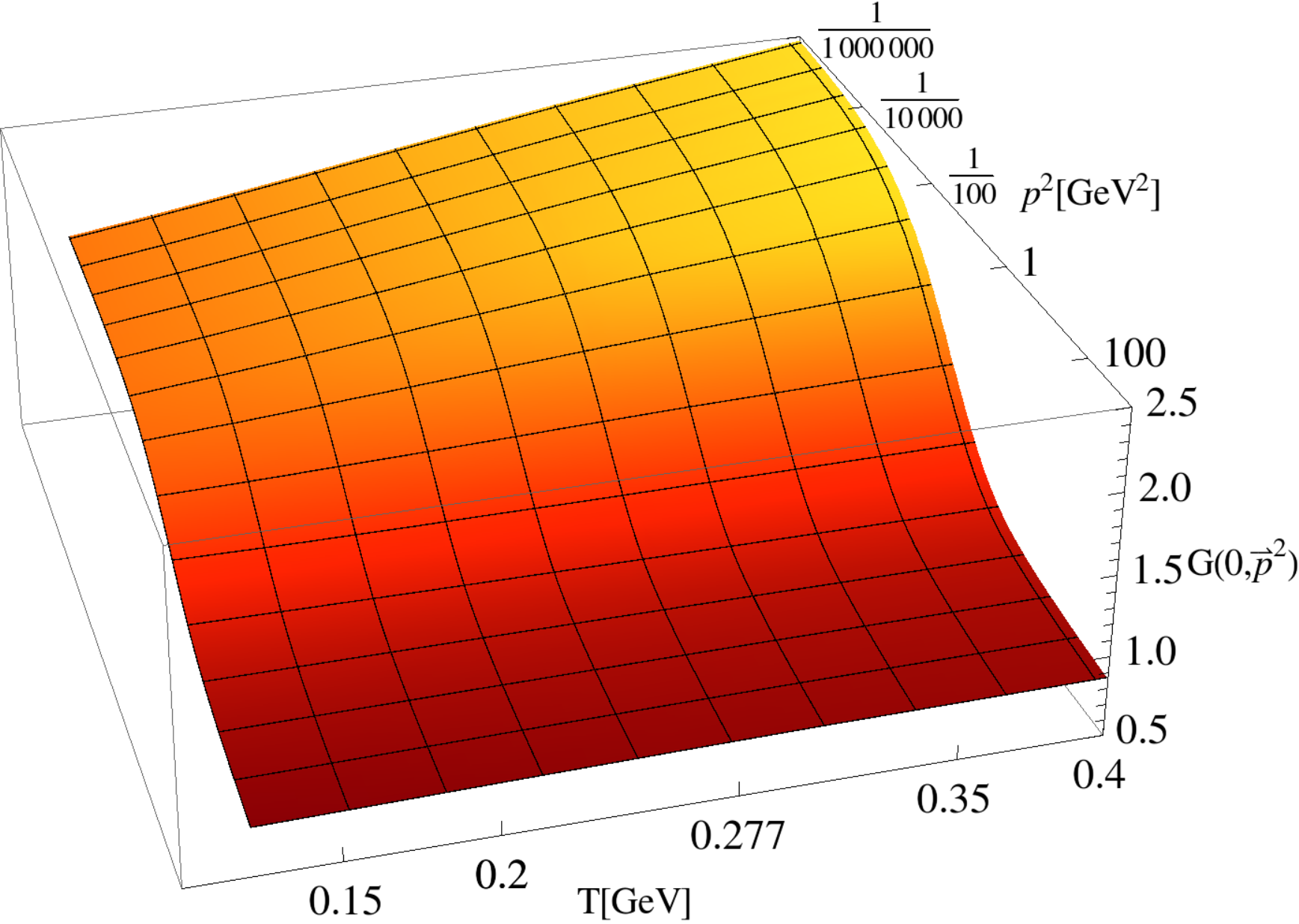}
\includegraphics[width=0.49\textwidth]{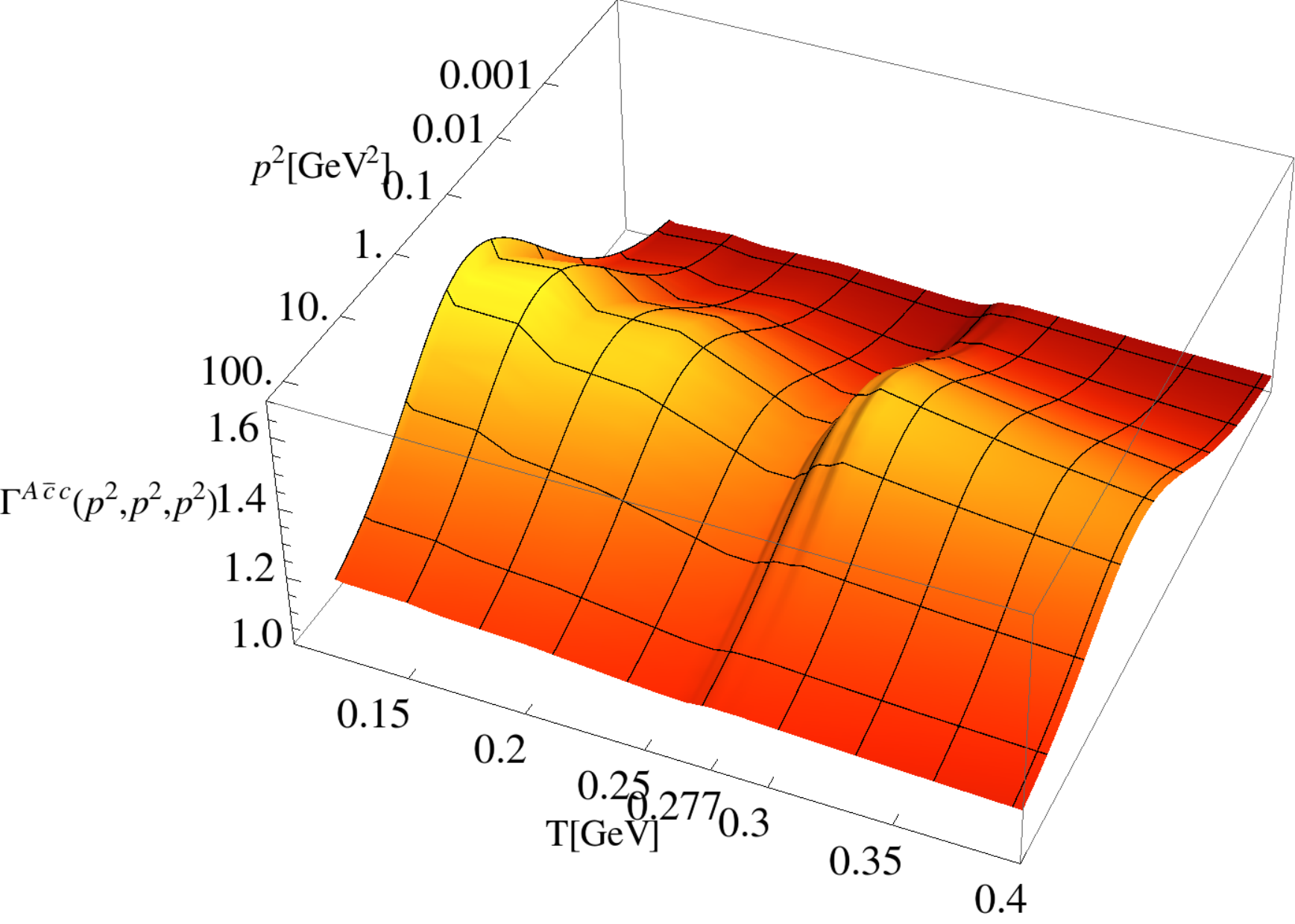}
\caption{\label{fig:gh-ghg_T}Ghost dressing function (left) and ghost-gluon vertex dressing function (right) at non-zero temperature.}
\end{center}
\end{figure}

At non-zero temperature fit functions for the gluon propagator dressings are available from ref.~\cite{Fischer:2010fx}. With DSEs the gluon propagator dressings were calculated in \cite{Maas:2005hs,Cucchieri:2007ta}. We used them for the calculation of the ghost dressing function. Using a bare ghost-gluon vertex we obtained the results shown in \fref{fig:3g_gh_T}. Despite the approximated ghost-gluon vertex good agreement with lattice data is seen. As expected from previous results from the lattice, e.g., \cite{Fischer:2010fx,Cucchieri:2007ta}, and flow equations \cite{Fister:2011uw}, we do not find any sign of the phase transition in the ghost propagator. From the functional point of view this can be understood directly from the corresponding DSE: At the zeroth Matsubara frequency, $p_0=0$, the loop with the chromoelectric gluon propagator reads explicitly:
\begin{align}
 -g^2 N_c \int_q \frac{
q_0^2\, \vec{k}\cdot\, \vec{p}\,(\vec{k}^2+\vec{k} \cdot \vec{q})
}
{
\vec{p}^2\,\vec{k}^2\,(q_0^2+\vec{k}^2)^2(q_0^2+\vec{q}^2)
}
G(\vec{q}^2,q_0) Z_L(\vec{k}^2, -q_0)
\end{align}
where $\vec{k}=\vec{p}-\vec{q}$. The sum over Matsubara frequencies does not have a contribution from $q_0=0$ due to the factor $q_0^2$ in the numerator. The first non-vanishing contribution stems from $q_0=2\pi\,T$. Since the chromoelectric gluon dressing $Z_L$ for non-zero Matsubara frequencies can be approximated by $Z_L(p_0, \vec{p}^2) \sim Z_L (0, p_0^2 + \vec{p}^2)$ \cite{Fischer:2010fx} and $p_0^2=1.74\,GeV$ at the phase transition, one can see that this integral is not sensitive to the phase transition. This approximation was employed for the gluon dressings only, while for the ghost propagator also higher frequencies were explicitly calculated. At temperatures above $200\,MeV$ deviations from an equivalent approximation were only found for $n=1$ and were within $5\,\%$. This is in agreement with \cite{Fischer:2010fx,Fister:2012th}. In \fref{fig:gh-ghg_T} the smooth transition of the ghost propagator dressing at $T_c$ can be seen.

The ghost propagator serves as a good test of the employed functional framework. As a next step we calculated the ghost-gluon vertex, for which to our knowledge currently no lattice data is available at non-zero temperature. Results from the functional renormalization group can be found in ref.~\cite{Fister:2011uw}. For this calculation we used the gluon propagator dressing fits and the results for the ghost propagator from above. At the symmetric point the temperature and momentum dependence is shown in \fref{fig:gh-ghg_T}. A small kink at $T_c$ is visible, but this may be an artifact of the truncation; for example, the dressed three-gluon vertex is not taken into account yet. The other wiggles in the dressing can directly be traced back to stem from the fits of the gluon propagator dressings.

\section{Summary}

At zero temperature we calculated the ghost and gluon propagators and the ghost-gluon vertex and compared them to lattice results. Using an optimized effective three-gluon vertex whose functional form is motivated by lattice results, we find good agreement. At non-zero temperature we took advantage of available lattice data for the gluon propagators. The calculation of the ghost dressing function served as a check for our setup. We also explained how one can understand from the ghost propagator DSE that the ghost is not sensitive to the phase transition although it couples to the chromoelectric gluon. Finally, we calculated the ghost-gluon vertex.

In general the calculation of vertices with functional methods has the advantage of being able to resolve the complete momentum dependence, whereas on the lattice typically specific momentum configurations are calculated. It should be stressed that such results nevertheless provide insight into the general behavior of Green functions and are useful for gauging the truncation dependence of functional results. Results for three-point functions at non-zero temperature can be used, for example, in the calculation of the Polyakov loop potential along the lines of \cite{Fister:2013bh,Fischer:2013eca}.

\section*{Acknowledgments}
This work was supported by the Helmholtz International Center for FAIR within the LOEWE program of the State of Hesse, the European Commission, FP7-PEOPLE-2009-RG No. 249203 and the Alexander von Humboldt foundation. 

\bibliographystyle{utphys_mod}
\bibliography{literature_YM4d_Lattice2013}

\end{document}